\documentclass{llncs}
\usepackage{llncsdoc}
\usepackage{
amsmath,
amssymb,verbatim,graphicx,color,layout}
\def\Ave{\operatorname{Ave}}

\def\bs{\backslash}

\def\FRR{\operatorname{\it FRR}}
\def\FAR{\operatorname{\it FAR}}
\def\AR{\operatorname{\it AR}}
\def\WAP{\operatorname{\it WAP}}

\def\HD{\operatorname{\it HD}}
\def\fHD{\operatorname{\it fHD}}

\def\match{\operatorname{\it match}}

\def\accept{\text{{\it accept}}}
\def\reject{\text{{\it reject}}}

\def\A{{\cal A}}
\def\U{{\cal U}}
\def\M{{\cal M}}

\def\disp{\displaystyle}

\def\iii{\text{\rm i\hspace{-.2ex}i\hspace{-.2ex}i}}
\makeatletter
\def\lapprox{\mathrel{\mathpalette\gl@align<}}
\def\gapprox{\mathrel{\mathpalette\gl@align>}}
\def\gl@align#1#2{\lower.6ex\vbox{\baselineskip\z@skip\lineskip\z@
\ialign{$\m@th#1\hfil##\hfil$\crcr#2\crcr\sim\crcr}}}
\makeatother
\newcommand{\tabtopsp}[1]{\vbox{\vbox to#1{}\vbox to1zw{}}}  %  Œrü'Æ'ÌŠÔŠu'ð‹ó%
\begin{document}
\markboth{\LaTeXe{} Class for Lecture Notes in Computer
Science}{\LaTeXe{} Class for Lecture Notes in Computer Science}
\thispagestyle{empty}
\title{Theoretical framework for constructing matching algorithms 
in biometric authentication systems}

\author{Manabu Inuma$^{1,2}$ \and Akira Otsuka$^{1,2}$ \and Hideki Imai$^{1,2}$}

\institute{Research Center for Information Security (RCIS) \\
National Institute of Advanced Industrial Science and Technology (AIST) \\
Akihabara-Daibiru Room 1003, 1-18-13, Sotokanda, Chiyoda-ku \\
Tokyo 101-0021 JAPAN
\and
Department of Electrical, Electronic, and Communication Engineering \\
Faculty of Science and Engineering, Chuo University \\
1-13-27 Kasuga, Bunkyo-ku Tokyo 112-8551 JAPAN}

\maketitle
\thispagestyle{empty}
\pagestyle{empty}

%%%%%%%%%%%%%%%%%%%%%%%%%%%%%%%%%%%%%%%%%%%%%%%%%%%%%%%%%%%%%%%%%%%%%%%%%%%%%%%%
\begin{abstract}
In this paper, we propose a theoretical framework to construct 
matching algorithms for any biometric authentication systems. 
Conventional matching algorithms 
are not necessarily secure against strong intentional 
impersonation attacks such as wolf attacks. 
The wolf attack is an attempt to impersonate a genuine user 
by presenting a ``wolf'' 
to a biometric authentication system 
without the knowledge of a genuine user's biometric sample. 
A ``wolf'' is a sample 
which can be accepted as a match with multiple templates. 
The wolf attack probability ($\WAP$) is 
the maximum success probability of the wolf attack, 
which was proposed by Une, Otsuka, Imai 
as a measure for evaluating security 
of biometric authentication 
systems \cite{UOI}, \cite{UOI2}. 
We present a principle for construction 
of secure matching algorithms against the wolf attack 
for any biometric authentication systems. 
The ideal matching algorithm 
determines a threshold 
for each input value 
depending on the entropy 
of the probability distribution 
of the (Hamming) distances. 
Then we show that if the information about the probability distribution 
for each input value is perfectly given, 
then our matching algorithm is secure against the wolf attack. 
Our generalized matching algorithm 
gives a theoretical framework 
to construct secure matching algorithms. 
How lower $\WAP$ is achievable depends on 
how accurately the entropy is estimated. 
Then there is a trade-off between the efficiency and 
the achievable $\WAP$. 
Almost every conventional matching algorithm 
employs a fixed threshold and hence it 
can be regarded as an efficient 
but insecure instance 
of our theoretical framework. 
Daugman's algorithm proposed in \cite{Daugman2} 
can also be regarded as 
a non-optimal instance of our framework. 

\end{abstract}

%%%%%%%%%%%%%%%%%%%%%%%%%%%%%%%%%%%%%%%%%%%%%%%%%%%%%%%%%%%%%%%%%%%%%%%%%%%%%%%%
\section{Introduction}
Biometric authentication systems automatically 
identify or verify individuals 
by physiological or behavioral characteristics. 
They are used in various services such as the immigration control 
at an airport, the banking transactions at an ATM, 
the access control to restricted areas in a building, and so on. 
The increase in the need of 
biometric authentication systems makes 
it important to explicitly evaluate the security of them. 
%\\
%We focus on the security against the intentional impersonation attack 
%such as a brute-force attack, and a zero-effort attack, 
%an artefact attack. 
\\
The false acceptance rate ($\FAR$) 
(see the definition $(\ref{def:FAR})$ 
in Section \ref{subsect:FARandWAP}) 
is traditionally used as a security 
measure against the zero-effort impersonation attack. 
The zero-effort approach assumes that an attacker 
will present his/her own biometric data. 
But, it is clearly not a rational assumption, 
since an attacker attempting to impersonate a genuine user will 
try to present a biometric data of the genuine user or its imitation. 
\\
Ratha et al. approximately calculate the success probability of 
a brute-force attack in a typical fingerprint-minutiae matching algorithm 
\cite{RCB}. 
The brute-force approach assumes 
that an attacker blindly selects an input value. 
However, if an attacker has some information 
about the algorithm employed in the system, 
the attacker might be able to find a sample 
which shows high similarity to most of the templates. 
Such a biometric sample is called a {\bf wolf} (cf. \cite{ISO}). 
An attacker could impersonate a genuine user 
with much higher probability than $\FAR$ by presenting a wolf 
to a biometric authentication system. 
\\
With regard to the artefact attack, 
Matsumoto et al. showed that some biometric authentication systems 
often falsely accepts some artefacts \cite{MMYH}. 
Therefore 
we should assume that 
an attacker may find a special input value 
not only from biometric samples 
but also from non-biometric samples. 
Une, Otsuka, Imai extended the definition 
of a wolf to include a non-biometric input 
value and defined the {\bf wolf attack probability ($\WAP$)} 
(see Definition \ref{def:WAP}) \cite{UOI}, \cite{UOI2}. 
$\WAP$ can be regarded 
as the upper bound of the success probability of attacks 
without the knowledge of a genuine user's biometric sample. 
Une, Otsuka, Imai proposed that $\WAP$ can be used as a security measure 
to evaluate the lower bound of a security level 
in a biometric authentication system. 
\\
Our goal is to propose a theoretical framework 
to construct matching algorithms 
for biometric authentication systems. 
Almost every 
conventional matching algorithm 
employs a fixed threshold 
determined based on $\FAR$ and 
the false rejection rate ($\FRR$) 
(see the definition (\ref{dfn:FRR}) in Section \ref{subsect:FRR}). 
It is not necessarily secure 
against the wolf attack. 
Une, Otsuka, Imai showed that 
in some of such matching algorithms, 
there actually exist 
strong wolves and $\WAP$ can be 
extremely higher 
\cite{UOI}, \cite{UOI2}. 
Surprisingly, as far as we know, 
no research have been conducted 
on security of matching algorithms 
until now. 
This is the first paper which 
studies the security characteristics 
of matching algorithms 
and gives a theoretical framework 
how to construct them securely. 
\\
Suppose a matching algorithm 
employs a threshold 
determined 
by the entropy of the probability distribution for each input value. 
We prove that if 
the entropy for each input value 
is perfectly given, 
then the matching algorithm 
is secure against the wolf attack 
(Theorem $\ref{thm:general}$, $\ref{thm:WAPcontrol}$). 
\\
In the real world, it might be difficult 
to perfectly calculate the entropy 
for each input value, however, 
a more accurate computation of the entropy 
can achieve a lower $\WAP$. 
Then there is a trade-off between the efficiency 
of the matching algorithm 
and the achievable $\WAP$ in the matching algorithm. 
\\
Previous results can be regarded as instances 
of our theoretical framework. 
Almost every previous matching algorithm employs 
a fixed threshold. 
In our theoretical framework, 
it can be regarded as an efficient instance 
which assumes all input values have the same entropy. 
However,
as mentioned above, 
it is not exactly secure against the wolf attack. 
\\
Daugman proposed a matching algorithm in which 
a threshold is determined for each match 
by taking account the number 
of bits available for comparison \cite{Daugman2}. 
His method can also be regarded as 
an instance of our framework, which 
assumes every bit of a sample independently and 
identically contributes 
to the entropy of the probability distribution. 
$\WAP$ in his algorithm can be relatively lower 
than that in an ordinary algorithm employing a fixed threshold. 
However, his matching algorithm 
is not necessarily secure against the wolf attack 
(see details in Section \ref{sect:varthdeponlength}), 
since we have to assume that an attacker 
knows more accurate information about 
the probability distributions. 
\\
This paper continues as follows. 
In Section \ref{sect:FARandWAP}, we will 
briefly introduce a typical model of biometric authentication systems and 
give explicit definitions of $\FAR$, $\WAP$, 
and security against the wolf attack. 
Our proposal can be easily adapted 
to all matching algorithms of all modalities that 
employ symmetric prametric functions such as the ordinary (Hamming) distance 
as the dissimilarity measure. 
%However, we omit the detail of the adaptation in this paper. 
%In Section \ref{sect:framework}, 
We will construct matching algorithms 
in the general case (Theorem $\ref{thm:general}$) and 
in the normal distribution case (Theorem $\ref{thm:WAPcontrol}$) and 
show that these matching algorithms are secure 
against the wolf attack. 
They give a theoretical framework to construct 
secure matching algorithms 
for any biometric authentication systems. 
In Section \ref{sect:varthdeponlength}, 
we will reconsider previous results 
in our theoretical framework. 
%In Section \ref{sect:conclusion}, we will summarize our results. 

\section{Model (Preliminaries)}
\label{sect:FARandWAP}
A biometric authentication system can be used for verification or 
identification of individuals. 
In verification, a user of the system claims to have a certain identity and 
the biometric system performs a one to one comparison between 
the offered biometric data and the template which is linked to the claimed identity. 
In identification, a one to all comparison is performed between the offered data and 
all available template stored in the database to reveal the identity of an individual. 
In this paper, we will discuss verification systems. 
\\
Let ${\cal U}$ be a set of all possible users of the biometric authentication system. 
Namely ${\cal U}$ is a set of all human individuals. 
For each user $u\in\U$, the identity of $u$ can be denoted by $u$, namely 
the identities of users can be identified with ${\cal U}$. 
Let ${\cal M}$ be a finite set with a symmetric prametric function 
$d:{\cal M}\times {\cal M}\to{\Bbb R}$, namely 
$d(x,y)=d(y,x)$, $d(x,y)\geq0$, $d(x,x)=0$ for all $x,y\in{\cal M}$. 
%Note that $d$ is not necessarily a distance such as a Hamming distance. 
\\
In an enrollment phase, for any user $u\in{\cal U}$, 
an acquisition device measures a biometric data of $u$. 
After processing the measurement data and extracting relevant features, 
the features are represented as an element $t_u$ of ${\cal M}$. 
Then the template $t_u$ of $u\in\U$ is stored in the database of the system. 
In a verification phase (matching phase) $\match$, 
a user $v\in{\cal U}$ claims an identity $w\in {\cal U}$ 
and a biometric measurement is acquired from $v$. 
This measurement is also transformed into an element $s$ of ${\cal M}$. 
A matching process compares $s$ with $t_w$ and 
$\match$ generates a message, $\accept$ or $\reject$, by a predetermined threshold 
$\tau\in{\Bbb R}_{\geq0}$ as follows:
\[
\match(v,w)=
\left\{
\begin{array}{ll}
\accept&\text{if $d(s,t_w)<\tau$}\\
\reject&\text{if $d(s,t_w)\geq\tau$}\quad.
\end{array}
\right.
\]
Each user $u\in \U$ enrolls and offers a certain biometric sample of $u$ 
in an enrollment phase and a verification phase, respectively. 
Therefore $\U$ can be regarded as a set of the biometric samples of users. 
For each biometric sample $u\in {\cal U}$, let $X_u$ be a random variable on $\M$ 
representing noisy versions of $u$, 
namely $P(X_u=s)$ denotes the probability that biometric data of $u$ 
will be transformed into $s\in\M$. 
Assume that the $X_u$, $u\in\U$, are independent. 
\subsection{The false rejection rate}\label{subsect:FRR}
The false rejection rate ($\FRR$) is the probability that 
a genuine user is rejected, namely it is defined by 
\begin{align}
\FRR&
=\mathop{\Ave}_{u\in\U}
P(\,\match(u,u)=\reject\,)\nonumber\\
&=\frac{1}{n}\sum_{u\in\U}\sum_{
(s,t)\in\M\times\M\atop
d(s,t)\geq\tau}\hskip-.3cm
P(X_u=s)P(X_u=t)\nonumber\\
&=
1-
\frac{1}{n}\sum_{u\in\U}\sum_{
(s,t)\in\M\times\M\atop
d(s,t)<\tau}
\hskip-.3cm
P(X_u=s)P(X_u=t)
\label{dfn:FRR}
\end{align}
where $n=\#\,\U$. 
For each user $u\in\U$, 
let $\FRR_u$ denote the probability that 
the user $u$ with the correct identity claim $u$ will be rejected. 
Namely, $\FRR_u$ is defined by
\begin{align}
\FRR_u
&=
\sum_{
(s,t)\in\M\times\M \atop
d(s,t)\geq\tau
}
P(X_u=s)
P(X_v=t)\nonumber\\
&=1-
\sum_{
(s,t)\in\M\times\M \atop
d(s,t)<\tau
}
P(X_u=s)
P(X_v=t)
\label{FRRu}
\quad.
\end{align}
It is easy to check that 
$\FRR=\disp\frac{1}{n}\mathop{\sum}_{u\in\U}\FRR_u$. 
\subsection{The false acceptance rate}
\label{subsect:FARandWAP}
The false acceptance rate $(\FAR)$ is the probability 
that an offer of a user with a wrong identity claim will be incorrectly accepted, 
namely $\FAR$ is defined by
\begin{align}
\label{def:FAR}
\FAR
&=\mathop{\Ave}_{
(u,v)\in{\U\times\U} \atop 
u\neq v
}
P(\,\match(u,v)=\accept\,)\nonumber\\
&=
\frac{1}{n(n-1)}
\sum_{
(u,v)\in{\U\times\U} \atop 
u\neq v}\sum_{
(s,t)\in\M\times\M\atop
d(s,t)<\tau
}
P(X_u=s)P(X_v=t)\quad.
\end{align}
The measure $\FAR$ is 
traditionally used to express a recognition accuracy 
of biometric systems. 
It is also used 
as a measure to evaluate the security of systems 
against the zero-effort impersonation attack. 
\\
The zero-effort approach assumes 
that an attacker attempting to impersonate a genuine user 
will present his/her own biometric data. 
This assumption is clearly so far from reality, 
since an attacker will try to present a biometric data 
of a genuine user or its imitation. 
\subsection{The wolf attack probability}
Une, Otsuka, Imai proposed a new security measure 
for biometric authentication systems \cite{UOI}, \cite{UOI2}. 
If an attacker can find an input value which 
matches many templates, 
then he succeed in impersonating 
a genuine user with a higher probability than $\FAR$ 
by presenting the input value to the biometric authentication system. 
Such an input value obtained from a biometric sample 
is called a {\bf wolf} by many authors (cf. \cite{ISO}). 
However, such an input value might be obtained 
not only from biometric samples but also from 
non-biometric samples. 
Matsumoto et al. show by experimentation that 
some artefacts can be falsely accepted 
in some biometric authentication systems \cite{MMYH}. 
\\
Considering these facts, 
we will extend the definition of a wolf as follows. 
\\
Let ${\cal A}$ be a set of all possible samples 
including non-biometric samples such as artefacts or synthetic samples. 
For each $w\in\A$, let $\FAR_w$ 
denote the probability 
that the sample $w$ with a wrong identity claim $v\neq w$ 
will be incorrectly accepted and let 
$\AR_{w}$ denote the probability 
that the sample $w$ with random claim will be accepted. 
Namely, $\FAR_w$ and $\AR_{w}$ are respectively defined by
\begin{align}
\FAR_w&=\mathop{\Ave}_{v\in \U\bs\{w\}}\,P(\match(w,v)=\accept)\nonumber\\
&=\frac{1}{\#\,\left(\U\bs\{w\}\right)}
\sum_{v\in\U\bs\{w\}}
\sum_{
(s,t)\in\M\times\M \atop
d(s,t)<\tau
}
P(X_w=s)P(X_v=t)\, ,\label{FARw}\\
AR_w&=\mathop{\Ave}_{v\in\U}\,P(\match(w,v)=\accept)\nonumber\\
&=\frac{1}{n}
\sum_{v\in\U}
\sum_{
(s,t)\in\M\times\M \atop
d(s,t)<\tau
}
P(X_w=s)P(X_v=t)\label{ARw}\quad.
\end{align}
It is easy to check that 
$\FAR=\disp\frac{1}{n}\mathop{\sum}_{u\in\U}\FAR_u$. 
The following theorem describes the relation between 
$\FRR_w$, $\FAR_w$ and $\AR_w$. 
\begin{lemma}\label{lemma:ineqFAR} 
\begin{align}
\AR_w&=
\left\{
\begin{array}{ll}
\FAR_w&\text{if $w\in\A\bs\U$}\\
\disp\frac{1}{n}(1-\FRR_w)+\left(1-\frac{1}{n}\right)\FAR_w&\text{if $w\in\U$}\;.
\end{array}
\right.\label{eqn1}
\end{align}
Therefore it immediately follows that 
\begin{align}
\label{eqn2}
\frac{1}{n}\sum_{u\in\U}\AR_u=
\frac{1}{n}(1-\FRR)+\left(1-\frac{1}{n}\right)\FAR\;.
\end{align}
\end{lemma}
\begin{proof} 
For any $w\in \A\bs\U$, it is clear that $\AR_w=\FAR_w$, since $\U\bs\{w\}=\U$. 
For any $w\in\U$, 
from the definitions $(\ref{FRRu})$, $(\ref{FARw})$ and $(\ref{ARw})$ of 
$\FRR_w$, $\FAR_w$ and $\AR_w$, respectively, we have 
\begin{align*}
\AR_w&=
\frac{1}{n}
\sum_{
(s,t)\in{\M\times\M}\atop
d(s,t)<\tau
}P(X_w=s)P(X_w=t)\\
&\quad+
\frac{n-1}{n}\cdot\frac{1}{n-1}
\sum_{v\in\U\bs\{w\}}
\sum_{(s,t)\in{\M\times\M}\atop
d(s,t)<\tau}
P(X_w=s)P(X_v=t)\\
&=\frac{1}{n}(1-\FRR_w)+\left(1-\frac{1}{n}\right)\FAR_w\quad.
\end{align*}
Therefore the results follow. 
\qed
\end{proof}
Put 
\[
\AR=\disp\frac{1}{n}\sum\limits_{u\in\U}\AR_u=
\frac{1}{n}(1-\FRR)+\left(1-\frac{1}{n}\right)\FAR\quad.
\]
Note that usual biometric authentication systems obviously 
satisfy $\FAR\leq1-\FRR$, namely $\FAR\leq\AR$. 
\begin{definition}\label{def:wolf} 
$(${\rm cf. \cite[Definition 3]{UOI}}$)$
A {\bf wolf} is defined 
as a sample $w\in \A$ such that $\AR_w>\AR$. 
\end{definition}
For any $\AR< p\leq 1$, a wolf $w$ such that $\AR_w=p$ 
is called a $p$-{\bf wolf}. 
In particular, $1$-wolf is called a {\bf universal wolf}. 
\begin{definition}
\label{def:wolfattack}
{\rm \cite[Definition 4]{UOI}}
Assume the following two conditions. 
\begin{itemize}
\item[{\rm (i)}] The attacker has no information of a biometric sample 
of a genuine user to be impersonated. 
\item[{\rm (ii)}] The attacker has complete information 
of the algorithms employed in the enrollment phase and the verification phase. 
\end{itemize}
The {\bf wolf attack} is defined as an attempt 
to impersonate a genuine user 
by presenting $p$-wolves with large $p$'s 
to minimize the complexity of the impersonation attack. 
\end{definition}
\begin{definition}
[Wolf attack probability ($\WAP$)]
\label{def:WAP}
$(${\rm cf. \cite[Definition 5]{UOI}}$)$\quad
The {\bf Wolf attack probability} is defined by 
\begin{align}\label{eqn:WAP}
\WAP&=
\mathop{\max}_{w\in \A}
\mathop{\Ave}_{v\in\U}
\,P(\,\match(w,v)=\accept\,)=
\mathop{\max}_{w\in \A}\AR_w
\;.
\end{align}
\end{definition}
It is clear that $\AR\leq\WAP$. 
Therefore, if $\FAR\leq1-\FRR$, then we have 
$\FAR\leq\AR=\mathop{\Ave}\limits_{u\in\U}\AR_u\leq\WAP$. 
\begin{definition}[Security against the wolf attack]
\label{dfn:optimal} For any $\delta>0$, 
a biometric authentication system is $\delta$-secure against the wolf attack 
if $\WAP<\delta$, namely there exist no wolf $w\in\A$ such that $\AR_w\geq\delta$. 
\end{definition}
If we use only $\FAR$ as a security measure 
against impersonation attacks, 
then we cannot explicitly evaluate the security against wolf attacks. 
Une, Otsuka, Imai proposed to evaluate the security level 
against the wolf attack 
by computing $\WAP$ \cite{UOI}, \cite{UOI2}. 

\section{Matching algorithms secure against the wolf attack}
\label{sect:matchingalgorithm}
For each $s\in\M$ and $x\in{\Bbb R}_{\geq0}$, 
the probability $P_s(x)$ that a template $t\in\M$ obtained from 
a biometric sample of a user will be at a distance less than $x$ from $s$ is defined by 
\begin{align}\label{FARs}
P_s(x)&=
\frac{1}{n}\sum_{v\in\U}
\sum_{
t\in \M \atop
d(s,t)<x
}\,P(X_v=t)\quad.
\end{align}
Then we have
\[
\AR_w=\sum_{s\in\M}P(X_w=s)P_s(\tau)\quad.
\]
\subsection{General case}
Fix $\delta>0$. Then we will construct a matching algorithm 
$\delta$-secure against the wolf attack as follows. 
\\
Almost every conventional matching algorithm employs a fixed threshold $\tau$ 
predetermined based on $\FRR$ and $\FAR$. 
However, we will employ a threshold $\tau_s$ determined 
for each element $s\in \M$ obtained from the sample $w\in\A$ 
offered in the verification phase. 
For each $s\in\M$, put
\[
\tau_s=\max\{x\in{\Bbb R}_{\geq 0}\,|\,P_s(x)<\delta\}\;.
\]
Note that 
a set $S=\{x\in{\Bbb R}_{\geq0}\,|\,P_s(x)<\delta\}$ 
is a non-empty closed subset of ${\Bbb R}_{\geq0}$ and therefore there exists 
the maximum of $S$. 
\\
For the implementation, we need to 
gather enough templates from each $v\in\U$ and estimate 
the probabilities $P(X_v=t)$ for all $t\in\M$. 
Then we can determine the threshold $\tau_s$ for each $s\in\M$ 
by doing the exhaustive search of all possible $x\geq 0$ such that $P_s(x)<\delta$. 
\\
It is clear that 
\begin{align*}
\WAP&=\mathop{\max}_{w\in \A}\sum_{s\in\M}P(X_w=s)P_s(\tau_s)<\delta\quad. 
\end{align*}
\\
The above discussion gives the following theorem. 
\begin{theorem}\label{thm:general}
If the information about the probability distribution 
$P_s(x)$ for each $s\in\M$ is completely given, 
then, for any $\delta>0$, 
we can construct a matching algorithm $\delta$-secure against the wolf attack. 
\end{theorem}

\subsection{Normal distribution case}
We assume that the distribution $P_s(x)$ is normal 
with mean $m_s$ and standard deviation $\sigma_s$ for each $s\in\M$, namely 
\begin{align}
\label{eqn:Gaussassump}
P_s(x)
&=
\disp\int_{-\infty}^{x}\frac{1}{\sqrt{2\pi}\sigma_s}
\exp\left(-\frac{1}{2}
\left(\frac{x-m_s}{\sigma_s}\right)^2
\right)\,dx
\end{align}
for any $x>0$. 
More strictly, we assume that $P_s(x)$ can be approximately estimated 
by the above equation. 
The distributions of Hamming distances 
for Daugman's iriscode satisfy this assumption (cf. \cite{Daugman1}, \cite{Daugman2}). 
Some authors use the Gaussian assumption as the basis of their analysis 
(cf. \cite{AYL1}, \cite{Ka1}, \cite{Wa1}). 
In general, the real-valued features will tend to approximate a Gaussian distribution 
when they are obtained by a linear combinations of many components, 
e.g. feature extraction techniques based on the principle component analysis (PCA) or 
the linear discriminant analysis (LDA) (cf. \cite{AYL1}). 
%This assumption will make our construction of an ideal matching algorithm simple 
%and optimal. 
\begin{comment}
In general the distribution of the templates obtained from the samples 
of different individuals If $d$ is the Hamming distance and the model of the system 
This assumption is valid, for example, in the case where 
$\M=\{0,1\}^l$ with $l$ large enough and 
almost all of the values $\sum\limits_{d(s,t)=x}P(X_u=t)$, $u\in\U$, 
are independent and identically distributed. 
\end{comment}
Under this assumption, we 
can construct a secure and simple matching algorithm and 
show that the matching algorithm is optimal, namely $\WAP$ is minimized to the (almost) 
same value as $\AR$. 
\\
Define the entropy $H(P)$ of the probability distribution $P$ by 
\[
H(P)=\int_{-\infty}^{\infty}-P(x)\log_2P(x)\;dx\;.
\]
By the assumption $(\ref{eqn:Gaussassump})$, 
it can be easily checked that 
\begin{align}\label{eqn:Gaussianentropy}
\disp H(P_s)=
\log_2\left(\sqrt{2\pi e}\cdot\sigma_s\right)\;.
\end{align}
We work with entropies $H(P)$ 
of continuous probability distributions $P$. 
Then the entropy $H(P)$ is not always non-negative. 
It is clear from $(\ref{eqn:Gaussianentropy})$ that 
if $\sigma_s<\disp\frac{1}{\sqrt{2\pi e}}$, 
then $H(P_s)<0$. 
Note that if a fixed threshold is employed, 
then an input value $s\in \M$ which 
has higher entropy $H(P_s)$ and therefore larger deviation 
$\sigma_s$ can be accepted with higher probability. 
\\
Fix a real number $\alpha$. 
For each $s\in\M$, put 
\begin{align}\label{dfn:varth}
\tau_s 
&=\alpha\sigma_s+m_s
=\disp\frac{\alpha2^{H_s}}{\sqrt{2\pi e}}+m_s
\end{align}
where $H_s=H(P_s)$. 
By the assumption $(\ref{eqn:Gaussassump})$, we have
\begin{align}
P_s(\tau_s)&=
\int_{-\infty}^{\tau_s}
\disp\frac{1}{\sqrt{2\pi}\sigma_s}
\exp\left(-\frac{1}{2}\left(\frac{x-m_s}{\sigma_s}\right)^2\right)\;dx
=\int_{-\infty}^{\alpha}
\disp\frac{1}{\sqrt{2\pi}}
\exp\left(-\frac{z^2}{2}\right)\;dz\label{const}
\end{align}
for all $s\in\M$. Put 
\[
\delta(\alpha)=\int_{-\infty}^{\alpha}\frac{1}{\sqrt{2\pi}}
\exp\left(-\frac{z^2}{2}\right)\;dz\quad.
\]
The following theorem can be immediately proved. 
\begin{theorem}\label{thm:WAPcontrol} 
Assume that the standard deviation $\sigma_s$ 
$($or the entropy $H_s$$)$ and the mean $m_s$ are 
perfectly given for each $s\in\M$. 
Then the matching algorithm employing 
the thresholds $\tau_s$, $s\in\M$, defined by $(\ref{dfn:varth})$ is 
$\delta(\alpha)$-secure against wolf attacks. 
Moreover, we have $\AR_w=\AR=\WAP=\delta(\alpha)$ for all $w\in\A$. 
\end{theorem}
\begin{proof} By the calculation $(\ref{const})$, for all $w\in\A$, we have 
\[
\AR_w=\sum_{s\in\M}P(X_w=s)P_s(\tau_s)=\delta(\alpha)\;.
\]
Therefore the results follow.
\qed
\end{proof}
Our generalized matching algorithm gives a theoretical framework 
for constructing matching algorithms secure 
against the wolf attack for any biometric authentication system. 
Under the ideal condition that for each $s\in \M$, 
the distribution $P_s(x)$ is completely calculated, 
our matching algorithm is optimal against the wolf attack. 
\\
In the real world, it might be difficult 
to explicitly calculate the distribution $P_s(x)$ 
for all $s\in\M$, however, 
a more accurate computation of 
$\sigma_s$, $H_s$, or $m_s$ 
for each $s\in\M$ can achieve a lower $\WAP$. 
Consequently, there is a trade-off between 
the efficiency of the matching algorithm 
and the security evaluated by the achievable $\WAP$. 
In the next section, we will reconsider 
previous results as instances of our theoretical framework. 
%
\begin{comment}
At the end of this section, 
we mention the influence of our proposed matching algorithm on the false 
rejections of genuine users. 
The purpose of our matching algorithm is 
to prevent the attacker 
from impersonating a genuine user by presenting an input value 
$s\in\M$ which shows high similality to most templates. 
Even a genuine user can be unfortunately rejected if he/she presents 
such a suspicious input value. 
However, a false rejection of such a badly behaved sample 
inevitably arises in a matching algorithm secure against the wolf attack. 
It is rather a critical security hole that 
a conventioal matching algorithm 
perhaps accepts such a badly behaved input value 
even though it is presented by an attacker. 
\end{comment}

\section{Previous results in our framework}
\label{sect:varthdeponlength}
In this section, we will review previous results 
in the context of our theoretical framework. 
\\
A conventional matching algorithm 
employing a fixed threshold 
can be viewed as an efficient instance 
of our framework, 
which assumes 
every input value has 
a constant entropy 
instead of computing 
the entropy for each input value. 
Such a matching algorithm 
is not secure against the wolf attack. 
\\
Daugman proposes 
a matching algorithm which employs a variable threshold 
in place of a fixed threshold as follows \cite{Daugman2}. 
He employs a fractional Hamming distance $d=\fHD$ defined by 
$\fHD(s,t)=\disp\frac{\HD(s,t)}{k}$ for any $s,t\in\M=\{0,1\}^{2048}$, 
where $k$ is the number of bits available for comparison. 
He determines a threshold depending on $k$ as follows:
\begin{align}\label{Daugth}
\tau(s,t)=\frac{\alpha'}{\sqrt{k}}+\frac{1}{2}
\end{align}
where $\disp\frac{1}{2}$ is the average of $\fHD(s,t)$ 
estimated from his database. 
His algorithm can also be regarded as an instance 
of our framework, which assumes 
every bit of each sample independently and identically 
contributes to the probability distribution. 
\\
However, his algorithm is not necessarily secure against the wolf attack, 
since every bit is not exactly independent and identical and 
the distributions $P_s(x)$, $s\in\M$, 
can be considerably different from each other. 
We assume that an attacker has 
more accurate information about 
the distributions 
%$P_s^{\diff,n}$ and 
$P_s(x)$, $s\in\M$. 
If the attacker can successfully find a smart input value 
$s\in\M$ such that the entropy 
$H(P_s(x))$ is extremely high, 
then he can be incorrectly accepted 
with much higher probability 
than $\AR$. 
\\
Daugman's matching algorithm is not always secure 
against the wolf attack, however, it motivated us to 
research a theoretical framework to construct 
secure matching algorithms. 
%His algorithm 
%can be viewed as an interesting instance of our theoretical framework. 
%
%

\end{document}